\documentclass[3p,times,procedia]{elsarticle}
\usepackage{nupha_ecrc}

\volume{00}

\firstpage{1}

\journalname{Nuclear Physics A}

\runauth{}

\jid{nupha}

\jnltitlelogo{Nuclear Physics A}

\usepackage{amssymb}

\usepackage{lineno}

\usepackage[figuresright]{rotating}

\usepackage{graphicx}
\usepackage{multirow}
\usepackage{epstopdf}
\usepackage{caption}
\usepackage{color}
\usepackage{bm}
\usepackage{textcomp}

\begin{document}

\begin{frontmatter}



\dochead{XXVIIth International Conference on Ultrarelativistic Nucleus-Nucleus Collisions\\ (Quark Matter 2018)}

\title{Search for the Chiral Magnetic Wave with Anisotropic Flow of Identified Particles at RHIC-STAR}


\author{Qi-Ye Shou (for the STAR Collaboration)}

\address{Shanghai Institute of Applied Physics, Chinese Academy of Sciences, Shanghai 201800, China}

\begin{abstract}
The chiral magnetic wave (CMW) has been theorized to propagate in the Quark-Gluon Plasma formed in high-energy heavy-ion collisions. It could cause a finite electric quadrupole moment of the collision system, and may be observed as a dependence of elliptic flow, $v_{2}$, on the asymmetry between positively and negatively charged hadrons, $A_{\rm ch}$. However, non-CMW mechanisms, such as local charge conservation (LCC) and hydrodynamics with isospin effect, could also contribute to the experimental observations. Here we present the STAR measurements of elliptic flow $v_{2}$ and triangular flow $v_{3}$ of charged pions, along with $v_{2}$ of charged kaons and protons, as functions of $A_{\rm ch}$ in Au+Au collisions at $\sqrt{s_{\rm NN}}$ = 200 GeV. The slope parameters of $\Delta v_{2}$($A_{\rm ch}$) and $\Delta v_{3}$($A_{\rm ch}$) are reported and compared to investigate the LCC background. The similarity between pion and kaon slopes suggests that the hydrodynamics is not the dominant mechanism. The difference between the normalized $\Delta v_{2}$ and $\Delta v_{3}$ slopes, together with the small slopes in p+Au and d+Au collisions at $\sqrt{s_{\rm NN}}$ = 200 GeV, suggest that the CMW picture remains a viable interpretation at RHIC.
\end{abstract}


\end{frontmatter}




\section{Introduction} 
It is proposed that, in a chirally symmetric phase, the chiral magnetic effect (CME) and the chiral separation effect (CSE) could intertwine to form a collective excitation of the Quark-Gluon Plasma (QGP), known as the chiral magnetic wave (CMW), a long wavelength hydrodynamic mode of chirality charge densities~\cite{Burnier2011,Kharzeev2016}. CMW is a signature of the chiral symmetry restoration, and manifests itself as a finite electric quadrupole moment of the collision system, where the "poles" ("equator") of the produced fireball acquire additional positive (negative) charge. This effect, if present, may lead to charge-dependent elliptic flow, $v_2$, i.e., the difference of $v_2$ between negatively and positively charged particles, $\Delta v_2$, could be proportional to the charge asymmetry, $A_{\rm ch}$, defined as $A_{\rm ch} = (N_{+} -N_{-}) / (N_{+} +N_{-})$ with {$N_{+}$} ({$N_{-}$}) denoting the number of positive (negative) particles in a given event.

Such charge asymmetry dependence of $v_2$ has been observed in Au+Au collisions in the STAR experiment~\cite{Adamczyk2015} and in Pb+Pb collisions in the ALICE experiment~\cite{Adam2016}. However, non-CMW mechanisms could also contribute to such linear relationship between $A_{\rm ch}$ and $v_2$. A hydrodynamic study~\cite{Hatta2016} claims that the simple transport of charges in QGP, convoluted with certain initial conditions related to isospin chemical potential, can lead to a sizable $v_2$ splitting of $\pi^\pm$. This model further predicts negative $\Delta v_{2}(A_{\rm ch})$ slopes for charged kaons and protons with larger magnitudes than the pion slopes. On the other hand, local charge conservation (LCC), together with the characteristic shape of $v_2(\eta)$ and $v_2(p_T)$, could also qualitatively explain the data~\cite{Bzdak2013}. In particular, it is predicted that such mechanism could give a similar effect on $v_3$, with the slope of $\Delta v_{3}(A_{\rm ch})$ being consistent with that of $\Delta v_{2}(A_{\rm ch})$ after normalization. In addition, in small system collisions, e.g., p+A, the magnetic field direction is presumably decoupled from the reaction plane~\cite{Relmont2017}, therefore the $\Delta v_{2}$ slopes, if observed, should be free of CMW signal and are dominated by background contributions such as LCC. These measurements have been performed by the CMS experiment~\cite{Sirunyan2017} in p+Pb and Pb+Pb collisions at $\sqrt{s_{\rm NN}}$ = 5.02 TeV, which is consistent with the LCC picture. However, one should be aware that the magnetic field strength decays in the vacuum much faster at the LHC energies than at RHIC, and by the time of quark production, the magnetic field may become too weak to initiate the CMW~\cite{Kharzeev2016,Koch2017,Deng2012}. The potential difference in the physics mechanisms between RHIC and the LHC motivates us to measure the $A_{\rm ch}$ dependence of anisotropic flow for various particle species using the STAR experiment at RHIC energy. 

Here we present the results of the STAR experiment for (1) dependence of the $\Delta v_{2} (A_{\rm ch})$ slopes on centrality for kaons and (anti)protons in Au+Au collisions at $\sqrt{s_{\rm NN}}$ = 200~GeV, which tests the hydrodynamics scenario; (2) dependence of the $\Delta v_{3} (A_{\rm ch})$ slope on centrality for $\pi^{\pm}$ in Au+Au collisions at $\sqrt{s_{\rm NN}}$ = 200~GeV, which studies the contribution from LCC; (3) the $\Delta v_{2} (A_{\rm ch})$ slopes in p+Au, d+Au and U+U collisions, which provide more information and further helps to disentangle the CMW signal from the background.    


\section{Data selection and analysis method}
A general overview of STAR experiment and detector can be found in~\cite{Arsene2005}. For this analysis, the large-system data sample consists of minimum-bias triggered events taken by the STAR detector from 2010 to 2016, including Au+Au, p+Au and d+Au collisions at $\sqrt{s_{\rm NN}}$ = 200 GeV, as well as U+U collisions at $\sqrt{s_{\rm NN}}$ = 193 GeV. The primary vertex of each event is required to be within 30 cm from the detector center along the beam direction, and within a radius of 2 cm from the beam along the transverse direction. All tracks are required to be within one unit around midrapidity ($|\eta| < 1$). The charged asymmetry $A_{\rm ch}$ is calculated for charged particles within transverse momentum window, $0.15 < p_T < 12$ GeV/c, excluding the low $p_T$ (anti)protons ($p_T < 0.4$ GeV/c) which are potentially contaminated by beam-pipe knockouts. For a given centrality, events are divided into five sub-groups according to their $A_{\rm ch}$, with each subgroup having similar number of events. The true $A_{\rm ch}$ is calculated by correcting the measured $A_{\rm ch}$ for detector efficiency obtained with HIJING and GEANT Monte-Carlo simulation of the STAR detector response.

To measure the $v_n$, the distance of the closest approach to primary vertex of the tracks is required to be less than 1 cm to reject secondary particles, and the transverse momentum is required to be within $0.15 < p_T < 0.5$ GeV/c to guarantee good particle identification and an almost constant mean $p_T$ as a function of $A_{\rm ch}$. Time projection chamber (TPC) and time-of-flight (TOF) detectors are jointly used to identify particle species, such as $\pi$, $K$ and $p$. Most technical details of the analysis are kept to be the same as the previous STAR publication~\cite{Adamczyk2015}.

The Q-cumulant method~\cite{Bilandzic2011} is adopted to extract the anisotropic flow. In this approach, all multiparticle cumulants are expressed with respect to flow vectors, $Q_n\equiv\sum_{k=1}^Me^{in\varphi_k}$. For instance, two-particle correlations for a single event and for all events, respectively, can be calculated by:
\begin{center}
{$\langle 2^{'} \rangle = \frac{p_{n} Q^{*}_{n}-m_{q}}{m_{p}M-m_{q}}$},~~~~~~
{$\langle\langle 2^{'} \rangle\rangle = \frac{\sum^{N}_{i=1}(w_{\langle 2^{'} \rangle})_{i}\langle 2^{'} \rangle_{i}}{\sum^{N}_{i=1}(w_{\langle 2^{'} \rangle})_{i}}$},
\end{center}
where {$p_{n}$} and {$Q_{n}^{*}$} are flow vectors and $w_{\langle 2^{'} \rangle}$ represents the event weight. The {$m_{p}$} and $M$ are the number of particles of interest (POI) and the number of reference particles (RFP), respectively, while {$m_{q}$} denotes the number of particles labeled by both POI and RFP. Using differential second-order cumulant, $d_{n}\{2\}=\langle\langle 2^{'} \rangle\rangle$, one can estimate differential flow by $v_{n} \{2\} = d_{n}\{2\} / \sqrt{c_{n}\{2\}}$, where $c_{n}$ represents the reference flow. An {$\eta$} gap of 0.3 is applied between POI and RFP to suppress short-range non-flow effects.


\section{Results} 
\subsection{Dependence of the $\Delta v_{2} (A_{\rm ch})$ slope on centrality for kaons and (anti)protons}

In addition to the aforementioned viscous hydrodynamic model with certain assumptions on isospin asymmetry, which predicts a stronger $v_{2}$ splitting in reverse order for $K^\pm$ than $\pi^\pm$~\cite{Hatta2016}, kaons and protons could also behave differently than pions, owing to their larger differences in the absorption cross sections between particles and antiparticles in the hadronic stage~\cite{Burnier2011}. Moreover, the Chiral Vortical Effect (CVE)~\cite{Kharzeev2011} could also contribute to the proton slope along with the CMW. Hence, the measurements of $\Delta v_{2} (A_{\rm ch})$ slopes for kaons and protons provide the direct test for all of these physics scenarios.

\begin{figure}[h]
\centering
\begin{minipage}{.5\textwidth}
  \centering
  \includegraphics[width=.9\linewidth]{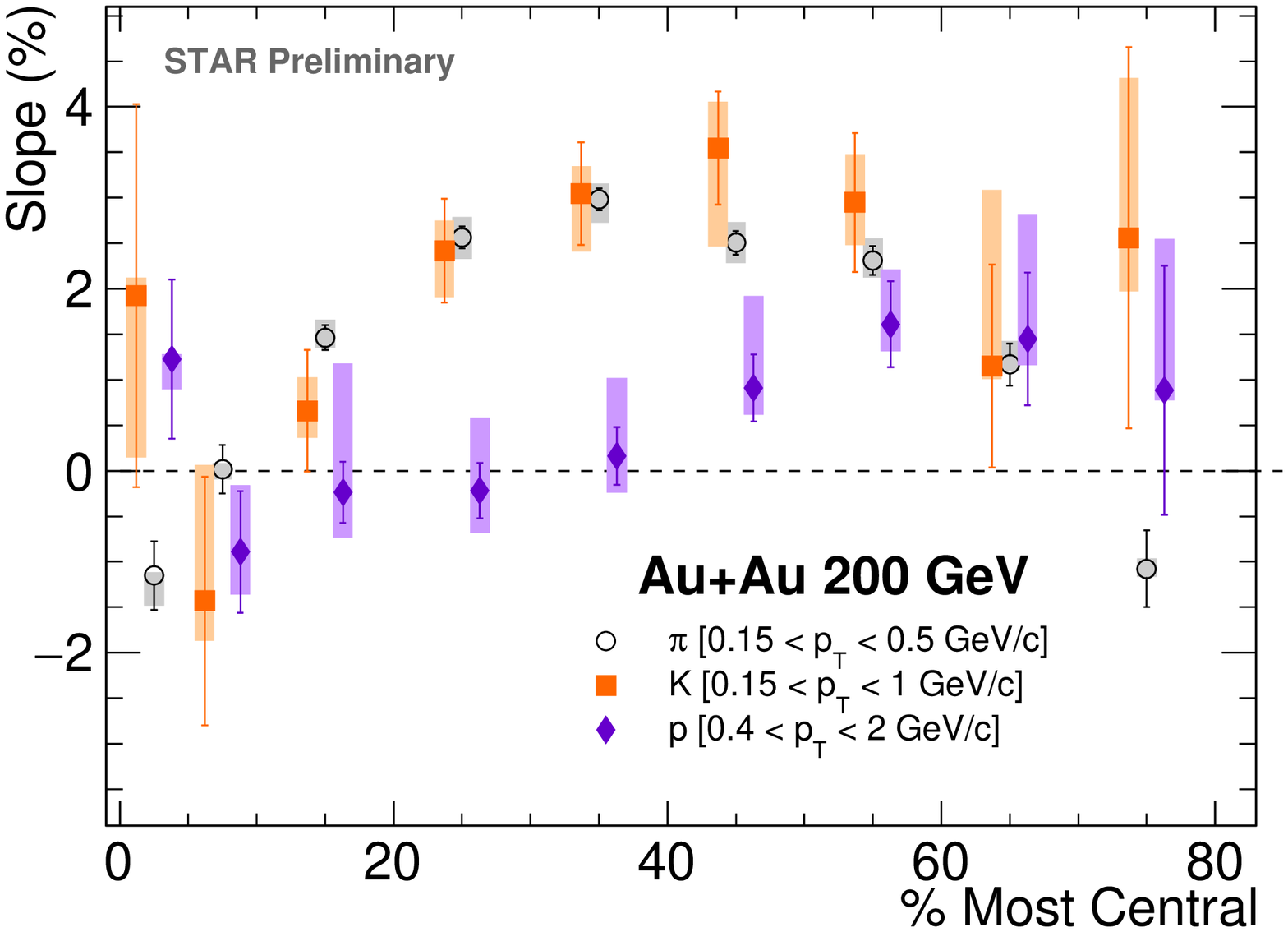}
  \captionof{figure}{Centrality dependence of the \label{label}$\Delta v_{2} (A_{\rm ch})$ slopes for kaons \\ and protons in Au+Au collisions at $\sqrt{s_{\rm NN}}$ = 200 GeV}
  \label{fig:slo_bes_k_p}
\end{minipage}%
\begin{minipage}{.5\textwidth}
  \centering
  \includegraphics[width=.9\linewidth]{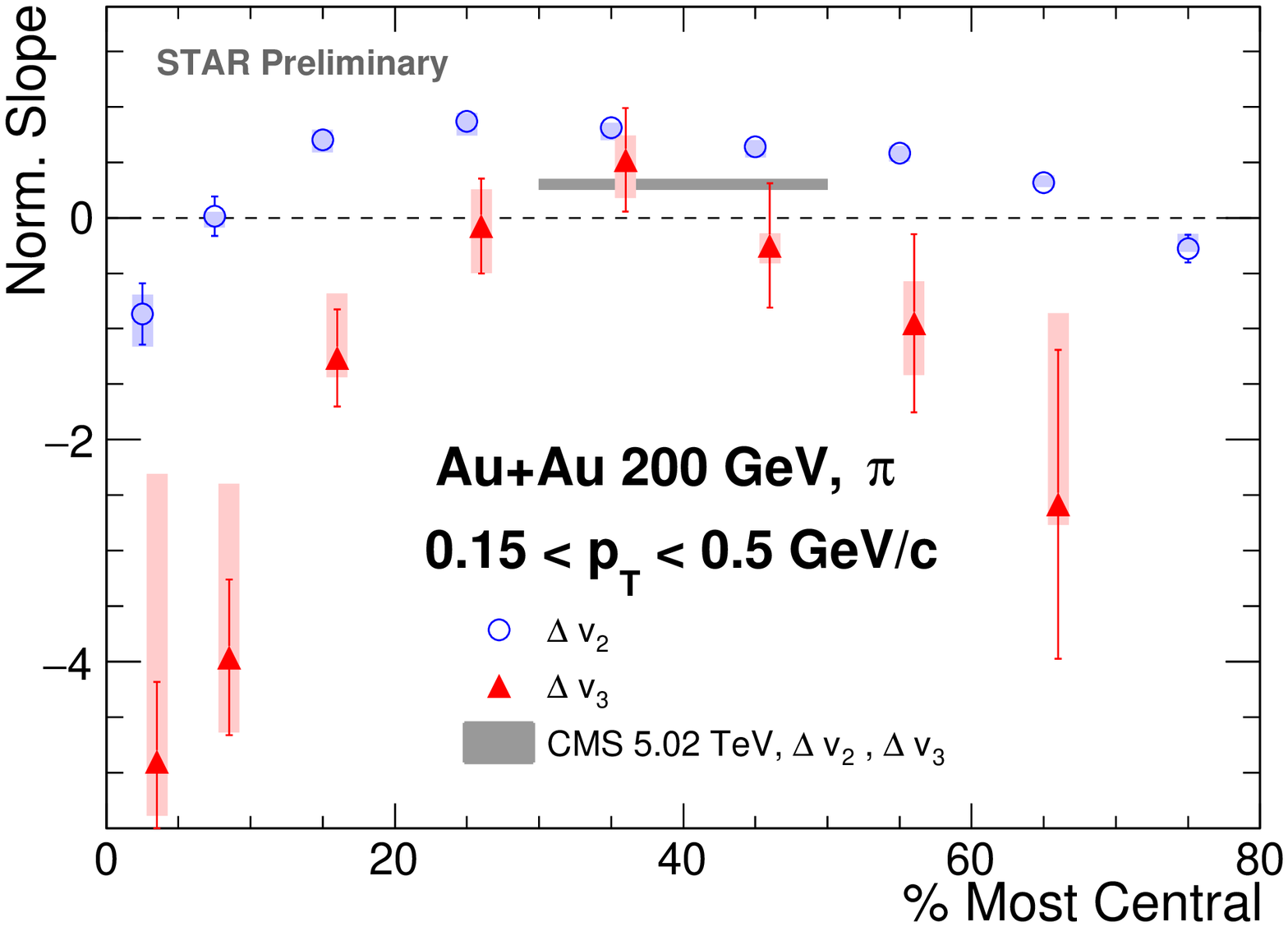}
  \captionof{figure}{The normalized $\Delta v_{2} (A_{\rm ch})$ and $\Delta v_{3} (A_{\rm ch})$ slopes for pions vs. centrality in Au+Au collisions at $\sqrt{s_{\rm NN}}$ = 200 GeV}
  \label{fig:v2v3}
\end{minipage}
\end{figure}

Fig.~\ref{fig:slo_bes_k_p} shows the centrality dependence of the kaon and proton slopes in Au+Au collisions at $\sqrt{s_{\rm NN}}$ = 200~GeV. The kaon slope displays a rise-and-fall trend with positive values in most centralities, highly consistent with the pion slope. It neither reveals a significant absorption effect, nor suggests that the hydrodynamics with isospin asymmetry is the dominant mechanism. It is also observed that the centrality dependence of proton\footnote{The $p_{T}$ coverage for (anti)proton is $(0.4, 2)$ GeV/$c$ for the sake of statistics} $\Delta v_{2}(A_{\rm ch})$ slopes are close to zero except for the centrality range $40-70\%$. The smaller proton slopes may suggest a mixed scenario without a dominating mechanism.

\subsection{Dependence of the $\Delta v_{3} (A_{\rm ch})$ slope on centrality for $\pi^{\pm}$}

Both the LCC effect and the viscous hydrodynamics with isospin asymmetry predict a linear dependence of $\Delta v_{3}$ on $A_{\rm ch}$ for pions, similar to that of $\Delta v_{2}$, while the electric quadrupole due to the CMW has no effect on $v_3$. Therefore, the $\Delta v_{3}(A_{\rm ch})$ slope, when properly normalized, $\Delta v_{n}^{norm.} = 2(v_{n}^{-} - v_{n}^{+})/(v_{n}^{-} + v_{n}^{+})$, provides an estimation of the background contribution to the $\Delta v_{2}(A_{\rm ch})$ slope.

Fig.~\ref{fig:v2v3} compares the normalized $\Delta v_{3} (A_{\rm ch})$ and $\Delta v_{2} (A_{\rm ch})$ slopes for pions as a function of centrality in 200 GeV Au+Au collisions. The normalized $\Delta v_{3} (A_{\rm ch})$ slopes for $0.15<p_T<0.5$ GeV/$c$ are lower than or consistent with zero for most centrality intervals, and are systematically below the normalized $\Delta v_{2} (A_{\rm ch})$ slopes. It is noticed that in semi-central collisions, the normalized slopes between $\Delta v_{2} (A_{\rm ch})$, $\Delta v_{3} (A_{\rm ch})$ and CMS results~\cite{Sirunyan2017} are consistent with each other. However, the large discrepancies in more central and the more peripheral collisions, where the $\Delta v_{3}$ slopes tend to go negative, suggest that the STAR measurements of the $\Delta v_{2}(A_{\rm ch})$ slopes for pions may not be purely dominated by the LCC effect.

\subsection{The $\Delta v_{2} (A_{\rm ch})$ slopes for $\pi^\pm$ in p+Au, d+Au and U+U collisions}

In small collision systems such as p+Au and d+Au, the orientation of the magnetic field is presumably decoupled from the 2$^{nd}$-order event plane~\cite{Relmont2017}, which makes such small systems an ideal testing ground for the observation of the disappearance of the $\Delta v_{2} (A_{\rm ch})$ slope. For p+Au and d+Au collisions at 200 GeV, the corresponding pion slopes are analyzed with the 2$^{nd}$-order event plane from TPC and are tested as a function of $N_{\rm part}$, the number of participating nucleons. The preliminary results of $\Delta v_{2} (A_{\rm ch})$ slopes in both p+Au and d+Au are consistent with zero within the uncertainties\footnote{Will be presented in the forthcoming STAR publication}. The disappearance of the CMW signal in these small systems demonstrates the magnitude of the possible background in the measurement of the CMW at RHIC. The $\Delta v_{2} (A_{\rm ch})$ slopes in U+U collisions are slightly higher than the results in Au+Au collisions, which could be qualitatively explained by the CMW picture due to a stronger magnetic field in U+U collisions.


\section{Summary} 

The experimental evidence of the CMW via $\Delta v_{2} (A_{\rm ch})$ slope for pions, has been challenged by the isospin effect and the LCC effect. In this work, we present the $\Delta v_{2} (A_{\rm ch})$ slopes for low-$p_{T}$ kaons and protons in Au+Au collisions at $\sqrt{s_{\rm NN}}$ = 200 GeV. The similarity between pion and kaon slopes suggests that the isospin effect is not the dominant contribution to the slopes. The LCC background has been investigated via the normalized $\Delta v_{3} (A_{\rm ch})$ slopes, which are typically smaller than the normalized $\Delta v_{2} (A_{\rm ch})$ slopes. The measured slopes in the small systems are consistent with zero within the uncertainties as expected. These measurements indicate that the CMW picture stays a viable interpretation of the pion $\Delta v_{2} (A_{\rm ch})$ slopes at RHIC. 


\section*{Acknowledgements}
This work was supported by the National Natural Science Foundation of China under Grant No. 11605070 and the National Key Research and Development Program of China under Grant No. 2016YFE0100900.






\bibliographystyle{elsarticle-num}
\bibliography{<your-bib-database>}



\end{document}